\documentstyle[twocolumn,prl,aps]{revtex}
\begin{document}
\draft
\title{Phase of the Wilson line at high temperature in the standard model}
\author{Chris P. Korthals Altes,$^a$ Kimyeong Lee,$^b$ and Robert D.
Pisarski$^c$}
\address{
$^a$Centre Physique Theorique au CNRS, B.P. 907, Luminy,
F 13288 Marseille, France\\
$^b$Physics Department, Columbia University, New York, NY 10027, USA\\
$^c$Department of Physics, Brookhaven National Laboratory,
P.O. Box 5000, Upton, New York 11973-5000, USA}
\date{\today}
\maketitle
\begin{abstract}
We compute the effective potential for the phase of the Wilson line
at high temperature in the standard model to one loop order.
Besides the trivial
vacua, there are metastable states in the direction
of $U(1)$ hypercharge.  Assuming that the universe starts out in
such a metastable state at the Planck scale, it easily persists to
the time of the electroweak phase transition, which then proceeds
by an unusual mechanism.  All remnants
of the metastable state
evaporate about the time of the $QCD$ phase transition.
\end{abstract}
\pacs{PACS numbers: }
\begin{narrowtext}

In gauge theories at nonzero temperature, the phase of the
Wilson line is a nonlocal order
parameter which distinguishes inequivalent
states \cite{gpy}-\cite{smilga}.  The Wilson line is defined as
$\Omega(x) = {\cal P}\;exp(i g \int^{1/T}_0 A_0(\vec{x},\tau) d\tau)$,
where $A_\mu$ is the gauge field, $g$ is the coupling constant,
and ${\cal P}$ denotes path ordering; we work at a
temperature $T$ in the
imaginary time formalism, so the imaginary time
$\tau$ runs from $0$ to $1/T$.
The Wilson line is a unitary operator, $\Omega^\dagger \Omega = 1$,
so that the phase of its vacuum expectation value, $\langle \Omega \rangle$,
can in principle be nonzero.
This phase vanishes in the usual, perturbative vacuum, $A_0 = 0$,
but it seems possible to have new states
in which the phase of $\langle \Omega \rangle$ is nonzero.

In scalar field theories the potential is
featureless at high temperature, and only develops structure ---
such as metastable states --- at low
temperature.  What is so interesting about the effective potential
for the phase of the
Wilson line is that it has structure in the high temperature,
plasma phase of a gauge theory.
Indeed, usually the Wilson line can only have a nontrivial
phase at high temperature.  For example, if a theory confines at
zero temperature, at low temperature $\langle \Omega \rangle = 0$,
and then it is meaningless
of speak of $\langle \Omega \rangle$'s phase.

Without dynamical fermions, the $SU(3)$ theory of the strong interactions
has a global $Z(3)$ symmetry.  This symmetry shows up in
the effective potential for $\Omega$'s phase as three degenerate vacua,
$\Omega = 1$ and $\Omega = exp(\pm 2 \pi i/3)$.
Adding dynamical fermions
lifts the $Z(3)$ degeneracy, so that the trivial vacuum at
$\Omega = 1$ is an absolute minimum.
Dixit and Ogilvie \cite{do} noticed
that while they are no longer degenerate with the trivial vacuum,
the states at $\Omega = exp(\pm 2 \pi i/3)$ are metastable if the number
of flavors is less than eighteen.

Ignatius, Kajantie, and Rummukainen \cite{ikr} then developed
an ``alternate'' cosmology: assuming that
the universe starts out with $\Omega = exp(\pm 2 \pi i/3)$
at the Planck scale, what are the cosmological
effects of its decay to $\Omega = 1$?
Ref. \cite{ikr} find that this
tunneling happens at temperatures of order $10 \; TeV$,
before the electroweak
phase transition, with at best only indirect
consequences for cosmology.

In this Letter we study the effective potential for the
phase of the Wilson line
to one loop order at high temperature in the standard model.
Our basic point is that because left handed quarks (for example)
carry quantum numbers under each of the three gauge groups,
it is not enough to
consider the effective potential
for any {\it single} group, such as $SU(3)$ color, but only that
for the {\it full} gauge group, $U(1) \times SU(2) \times SU(3)$.
We find that the metastable
state points in $SU(3)$ are unstable in the direction of
$U(1)$ hypercharge.
There are other states, however,
which are metastable in all directions, and so are candidates
for an alternate cosmology.
If the universe fell into such a metastable state at the $GUT$ or
Planck scale,
we find that that the metastable
state easily persists to the temperature of the electroweak
phase transition, where it dramatically affects when and how the electroweak
phase transition occurs.
Details are given elsewhere \cite{sintra}.

At the outset we confess that it is very controversial whether or not there
are any physically realizable
states with a nonzero phase for the Wilson line
\cite{van}-\cite{smilga}.  Some of the controversy
arises because the partition function which corresponds to a nonzero
phase is one in which there is an imaginary
chemical potential for the total charge of a system
\cite{twoloop,bgkp}.  Since this chemical
potential is imaginary, the ``entropy'', as customarily defined,
can be negative, which in thermodynamics is
nonsense.  We believe that the resolution of this
paradox lies in the fact that what is being computed is not an entropy
{\it per se}, but some more general quantity which need not be positive.
We note that for the states which we consider,
$t_1$ and $t_2$ as defined below, the ``entropy''
{\it is} positive, ${\cal S}_{1 \; loop}(t_{1,2}) < 0$, so for us this
is not an immediate problem.
For now we leave these questions aside, with purely incendiary intent, to
show that if the states with nonzero phase exist, then
they could well matter for the early universe.

In a nonabelian theory $\Omega$ is a matrix, of which only the
eigenvalues are gauge invariant \cite{bel,bgkp}.
These eigenvalues lie in the space of the mutually commuting generators
of the group, which is the Cartan subalgebra.
To parametrize a state in which the phase of $\langle \Omega \rangle$
is nonzero, at one loop order
it suffices to calculate the effective potential for a
constant, background field, $A_0$,
which lies in the Cartan subalgebra \cite{footnote1}.
In the standard model we take:
$$
{\cal A}_0 \; = \; \frac{2 \pi T}{g_{st}}
\left(
\begin{array}{ccc}
q/3 + r/2 & 0                   & 0               \\
0                  & q/3 - r/2   & 0               \\
0                 & 0           & - 2 q/3 \\
\end{array}
\right) \; ,
$$
\begin{equation}
A_0 \; = \;
\frac{2 \pi T}{g} \;
\left(
\begin{array}{cc}
s/2 & 0  \\
0 & -s/2 \\
\end{array}
\right)  \;\; , \;\;
B_0 \; = \; \frac{2 \pi T}{g'} \, t \; .
\label{e1}
\end{equation}
${\cal A}_0$ is the field for $SU(3)$ color, with coupling
constant $g_{st}$,
$A_0$ is the field for weak $SU(2)$, with coupling $g$,
and $B_0$ is the field for hypercharge $U(1)$, with coupling $g'$.
The Cartan subalgebra of the standard model
is a four dimensional space, $(q,r,s,t)$.
For the $SU(3)$ field ${\cal A}_0$ there are two
coordinates: $q$, proportional to
the $SU(3)$ matrix $\lambda_8$, and
$r$, proportional to the $SU(3)$ matrix $\lambda_3$.
For the $SU(2)$ field $A_0$, there is just one coordinate,
$s$, proportional to the $SU(2)$ matrix $\sigma_3$.
Lastly, there is the coordinate $t$ for $U(1)$ hypercharge.

For (\ref{e1}) we define the Wilson line as
$\Omega(q,r,s,t)=exp\{i( g_{st} {\cal A}_0 + g A_0 + g' B_0/6 )/T\}$.
The contribution to the Wilson line for the $SU(2)$ and $SU(3)$
gauge fields are the standard expressions for the fundamental representation
of each group.  For the $U(1)$ of hypercharge, however,
we choose the Wilson line which corresponds to
the {\it smallest} hypercharge, which in the standard model is $+1/6$.
The smallest charge enters because that fixes the
Wilson line to be periodic in $t$ with period $6$; the larger
charges all have smaller periods which are integral divisors of $6$.

At one loop order the potential for the phase of the Wilson line
involves the function
$$
v(z) = [z]^2 (1-[z])^2 \;\;\; , \;\;\; [z] \; = \; |z|_{modulo \; 1} \; .
$$
The potential only depends on $z$ through $[z]$,
the absolute value of $z$ modulo $1$.
The absolute value of $z$ enters because
the overall system is charge neutral so the sign of $z$ doesn't
matter.  Modulo $1$ enters because if the Wilson line depends upon $z$ as
$exp(2 \pi i z)$, then $z \rightarrow z + 1$ is a symmetry.

In the standard model the total potential is
\begin{equation}
{\cal S}_{1 \; loop}(q,r,s,t)
= \int d^3 x  \; \pi^2 T^3 \left( - \, \frac{427}{360}
\; + \; {\cal V}(q,r,s,t)\right) \; .
\end{equation}
The constant $-427/360$ is the free energy of the standard
model to lowest order, as a sum of
ideal gases.  The potential for the phase of the Wilson line is
$$
{\cal V}(q,r,s,t) \; = \;
\frac{4}{3} \, v(s)
\; + \; \frac{4}{3} \left( v(r) + v\!\left(q \pm \frac{r}{2} \right) \right)
$$
$$
\; + \; \frac{2}{3} \, v\!\left( \pm \frac{s}{2} + \frac{t}{2} \right)
\; - \; 2 \, v\!\left( - t + \frac{1}{2} \right)
\; - \; 2 \, v\!\left(\pm \frac{s}{2} - \frac{t}{2} + \frac{1}{2}\right)
$$
$$
\; - \; 2 \, v_{\cal C}\!\left(q,r,-\frac{t}{3} \right)
\; - \; 2 \, v_{\cal C}\!\left(q,r,+\frac{2t}{3} \right)
$$
\begin{equation}
\; - \; 2 \, v_{\cal C}\!\left(q,r,\pm \frac{s}{2}+\frac{t}{6} \right)
\; + 30 \, v\!\left( \frac{1}{2} \right) \; .
\label{e3}
\end{equation}
In this expression $v(q \pm r/2)$ denotes
$v(q+ r/2) + v(q - r/2)$, {\it etc.}, while colored fields in the
fundamental representation generate the function
$$
v_{\cal C}(q,r,z) \; = \;
v\!\left( \frac{q}{3} + \frac{r}{2} + z + \frac{1}{2} \right)
$$
\begin{equation}
\; + \; v\!\left( \frac{q}{3} - \frac{r}{2} + z + \frac{1}{2} \right)
\; + \; v\!\left( - \frac{2 q}{3} + z + \frac{1}{2} \right)  \; .
\label{e2}
\end{equation}
The terms in (\ref{e3}) can be read off from the charge assignments
of the standard model.  We assume that there is a single isodoublet
Higgs field and three generations of quarks and
leptons.  We compute in the limit of very high temperature
where the mass of the Higgs field, and spontaneous
symmetry breaking, can be ignored.
The contributions in (\ref{e3}) are due to,
respectively ($Y=$ hypercharge):
the $SU(2)$ gauge bosons ($Y=0$),
the $SU(3)$ gluons ($Y=0$),
the Higgs field ($Y=+1/2$), right handed leptons
($Y=-1$), left handed leptons ($Y=-1/2$),
right handed down, strange, and bottom quarks ($Y=-1/3$)
right handed up, charm and top quarks ($Y=+2/3$),
and left handed quarks ($Y=+1/6$).  Lastly there is an overall constant to
fix ${\cal V}(0,0,0,0) =0$.
Since left handed quarks
transform nontrivially under each part of
$U(1) \times SU(2) \times SU(3)$, the full potential ${\cal V}(q,r,s,t)$
is not separable.

Metastable points in the potential for the phase of the Wilson line
arise when there are fields with different charges to play off against
one another.
For example, for the $\lambda_8$ direction in $SU(3)$, gluons in
the adjoint representation have charge $\pm 1$ times the coupling constant,
while quarks in the fundamental representation have charge $+1/3$ and
$-2/3$.  The gluon contribution to ${\cal V}(q,0,0,0)$
is periodic in $q$ with period $1$,
while the quark contributions are periodic in $q$
with periods of $3$ and $3/2$.
The balance between these different
periods produces (for less than eighteen flavors) two metastable
points, $q_+ = (1,0,0,0)$ and $q_- = (2,0,0,0)$
($\Omega(q_{\pm}) = exp(\pm 2\pi i /3)$), which are
locally stable in the directions of $q$ and $r$ \cite{do,ikr}.
In the full standard model we find that $q_+$ and $q_-$
are locally stable in the directions of $q$, $r$,
and $s$, but are {\it un}stable in the $t$ direction.
Thus they are not possible
candidates for an alternate cosmology.  The instability of $q_\pm$ is
not obvious, and due to the detailed (hyper)charge assignments of the
standard model.

In the standard model there are many different
hypercharges: $+1/6$, $-1/3$, $-1/2$, $+ 2/3$, and $-1$.
This produces an exceedingly rich structure for the effective potential
in the direction of $U(1)$ hypercharge, $t$.
This is illustrated in the plot of $V(t)={\cal V}(0,0,0,t)$ in fig. (1).
(Since $V(6-t)=V(t)$, in fig. (1) we only show $t:0 \rightarrow 3$.)
There are two metastable points:
$t_1 = (0,0,0,1.28751...)$ and $t_2=(0,0,0,1.74843...)$,
where $V(t_1) > V(t_2) > 0$.
These points are locally stable in {\it all} four directions, $q$, $r$,
$s$, and $t$.
(The point $(0,0,0,3)$, which from fig. (1) is locally stable in the
$t$ direction, is unstable in the $s$ direction.)
There are several other metastable points in the $(q,0,0,t)$ plane which
are equivalent to $t_1$ and $t_2$.
Like $t_1$ and $t_2$, these other metastable points persist to the
time of the electroweak phase transition, although afterwards their evolution
may differ.  We discuss $t_1$ and $t_2$ as illustrative examples.

Assuming that the universe starts out in $t_1$ or $t_2$,
the probability for the universe to decay
is computed by standard techniques \cite{false}.
Following the computation of the interface tension by
Bhattacharya {\it et al.} \cite{bgkp},
we compute the stationary point of a
three dimensional effective action, ${\cal S}_{eff}$, which
is the sum of the
bare action and the potential term, ${\cal S}_{1 \; loop}$.
As appropriate at high temperature, we look for the $O(3)$ symmetric
bounce with minimal action, assuming that the motion is entirely
along the $t$ direction, which
seems reasonable from symmetry reasons in the plots of the full potential.

Since there are two metastable points, there are three possible
tunneling paths.  These were found numerically, without recourse to
the thin wall approximation:
$$
{\cal S}_{eff}(t_2 \rightarrow 0) \sim
\frac{1816.}{(g')^3} \;\; , \;\;
{\cal S}_{eff}(t_1 \rightarrow 0) \sim
\frac{276.}{(g')^3}\;\; , \;\;
$$
\begin{equation}
{\cal S}_{eff}(t_1 \rightarrow t_2) \sim
\frac{82.4}{(g')^3} \; .
\label{e4}
\end{equation}

To estimate the tunneling temperature we use
a naive analysis, much simpler than that of
Ignatius, Kajantie, and Rummukainen \cite{ikr}.
We assume that the decay rate per unit
space time volume is $\Gamma \sim T^4 \, exp(-{\cal S}_{eff})$,
and that tunneling occurs when $\Gamma t_c^4 \sim 1$, where
$t_c$ is the cosmological time.  In a
radiation dominated era $t_c\sim m_{Planck}/T^2$, so
the tunneling temperature is
$ T_{tunneling}  \sim  m_{Planck} \;
exp(- {\cal S}_{eff}/4)$.
Choosing the hypercharge coupling $g' = .535$, as appropriate at the
GUT scale, we find that the most likely tunneling,
$t_1 \rightarrow t_2$,
happens for $T_{tunneling} \sim 10^{-39} GeV$,
well below the temperature for the electroweak phase transition.
The temperatures for the other tunnelings are even lower.

It is easy to see why the metastable states for
$U(1)$ hypercharge do not readily decay, while
the states with $SU(3)$ color, $q_\pm$, do
(ignoring their instability in the $t$ direction).
The action for the two tunnelings is about equal at the
$GUT$ scale: for $t_1 \rightarrow t_2$,
${\cal S}_{eff} \sim 82/(g')^3$,
while for $q_\pm \rightarrow 0$,
${\cal S}_{eff} \sim 92/(g_{st})^3$.
In the analysis of ref. \cite{ikr},
the tunneling for $SU(3)$ color occurs because
the strong coupling, $g_{st}$, increases as the temperature
decreases, and is large enough
by $T \sim 10 \, TeV$ to trigger tunneling.
In contrast, the hypercharge
coupling $g'$ only gets smaller with decreasing temperature, so
its running is clearly negligible, since that only acts to make
tunneling even more improbable.

In actuality the tunneling temperature
is irrelevant, since the electroweak phase transition takes place
well before then.
The electroweak phase transition is radically changed by the
presence of a nonzero phase for the Wilson line.
If the Higgs part of the lagrangian is
%
%
${\cal L}_{Higgs}=
|D_\mu \phi|^2 - \mu^2 |\phi|^2 + \lambda (|\phi|)^2$,
then for one component of the
scalar its effective mass squared is
$m_{eff}^2 = -\mu^2 + (g'B_0/2)^2 = -\mu^2 + (\pi T t)^2$.
When $t = t_{1,2}\neq 0$, we see that we can compute the
temperature of the electroweak phase transition
at {\it tree} level by setting $m_{eff}^2 = 0$ to obtain
$T_{ew} = \mu/(\pi t)$.  This value
for $T_{ew}$ is smaller than the usual estimate when $t=0$: then one
balances one loop thermal terms against the negative mass squared term at
tree level to obtain $T_{ew} \sim \mu/g$, {\it etc.}
At tree level the transition for $t \neq 0$ is of second order,
but it is well possible that one loop effects drive it first order.

For temperatures below $T_{ew}$ spontaneous symmetry breaking occurs,
and the $SU(2)$ and $U(1)$ fields mix in the usual fashion.  A condensate
for $B_0$ generates two condensates, one for the timelike components
of the massive $Z_0$, $Z_0^0$, and the other
for the massless photon.

As the electroweak phase transition proceeds,
the $Z_0^0$ condensate relaxes to zero.
To see this, consider ${\cal L}_{Higgs}$, ignoring
(for the purposes of discussion)
the mixing of the $Z_0$ and the photon.  Below $T_{ew}$ the scalar
field has a vacuum expectation value
$\phi_0 = \sqrt{(\mu^2 - (\pi T t)^2)/(2\lambda)}$; inserting
this value of $\phi_0$ back in the lagrangian, the scalar potential
is ${\cal L}_{Higgs}(t) = - (\mu^2 - (\pi T t)^2)^2/(4 \lambda)$.
Since the overall sign of ${\cal L}_{Higgs}(t)$
is negative, the potential is always minimized
by letting $t \rightarrow 0$.
This illustrates our general comment that usually the phase of
the Wilson line is only interesting at high temperature, in the plasma phase
of a gauge theory.

The process by which the $Z_0^0$ condensate relaxes to zero could
have striking consequences for the electroweak phase transition.  Under
charge conjugation the Wilson line $\Omega \rightarrow \Omega^\dagger$,
and so its phase changes sign.
Thus $Z_0^0 \rightarrow 0$ serves as a source of $C$ and (as the
weak interactions are chiral) $CP$ violation.
We suggest that $Z_0^0 \rightarrow 0$ {\it might} serve as a
source of $CP$ violation in generating an asymmetry in baryon number
at the electroweak phase transition \cite{ew}.
The magnitude of the $CP$ violation is of order one and so rather large,
$\sim \pi t_{1,2}$.  This is perhaps the most appealing feature of
the alternate cosmology.

What of the remaining condensate for the
timelike component of the photon field,
which we define as $A^{em}_0 = 2 \pi T u/e$?
($u = cos^2\theta_w \, t$, with $\theta_w \sim .5$
the Weinberg angle.)
At temperatures
above the $QCD$ phase transition the relevant gauge group
for the Wilson line is $U(1)_{em} \times SU(3)$, and is a function of
three variables, $q$, $r$, and $u$.
At such temperatures the effective potential for $u$ is nontrivial,
since the photon couples both
to fields with fractional electric charge,
such as quarks, and to fields with integral electric
charge, such as leptons, $W^\pm$, {\it etc.}
Assuming that all fields are massless,
we find \cite{sintra}
that this potential has a nontrivial metastable minimum for
$u = u_1 \sim 1.22727...$, which is locally stable in all
directions, $q$, $r$, and $u$.
Given the known value of the Weinberg angle,
if the theory starts out at $t_1$ or $t_2$ above
$T_{ew}$, it falls into $u_1$ below $T_{ew}$.  Like $t_1$ and $t_2$,
$u_1$ does not readily decay by tunneling.

This potential for the phase of the Wilson line changes once
the temperature falls below that of the $QCD$ phase transition.  Below
temperatures of, say, $150 MeV$, only color singlet states contribute
to the potential, and we can forget about the $SU(3)$ Wilson line.
There is still a potential for the phase of the
electromagnetic Wilson line,
but it only receives contributions from fields with integral charge.
This potential has no metastable points, and so the point
$u_1$ dynamically relaxes to a value equivalent to the trivial vacuum,
where the electromagnetic $\Omega =1$.

Obviously we leave many questions unanswered.  Most importantly,
do the metastable points in
the potential for the phase of the
Wilson line represent physically realizable
states?  What of the potential for the
phase of the Wilson line at even higher temperatures?
We have computed in $SU(5)$ \cite{sintra}, and find metastable states which
can fall into either $t_1$ or $t_2$ at temperatures below the $GUT$ scale.
What is the potential for the
phase of the Wilson line in the high temperature phase
of a string theory, as operative at temperatures above the Planck scale?
Is it flat, or does it have metastable points as well?

This work of K.L. is supported
by the Department of Energy, the NSF Presidential Young
Investigator program and the Alfred P. Sloan Foundation.
The work of R.D.P. is supported by a DOE grant at
Brookhaven National Laboratory (DE-AC02-76CH00016).

\begin{figure}
\caption{
The potential for the Wilson line in the direction of $U(1)$ hypercharge,
$V(t)={\cal V}(0,0,0,t)$.
}
\label{f1}
\end{figure}
\end{narrowtext}

\begin{references}
%
\bibitem{gpy}
D. J. Gross, R. D. Pisarski, and L. G. Yaffe, Rev. Mod. Phys.,
{\bf 53}, 43 (1981); N. Weiss, Phys. Rev. {\bf D24}, 475 (1981) and
{\bf D25}, 2667 (1982).
%
\bibitem{twoloop}
R. Anishetty, J. Phys. {\bf G10}, 439 (1984);
A. Roberge and N. Weiss, Nucl. Phys. {\bf B275} [FS17], 734 (1986);
V. M. Belyaev and V. L. Eletsky, Z. Phys. {\bf C45}, 355 (1990);
K. Enqvist and K. Kajantie, Z. Phys. {\bf C47}, 291 (1990).
%
\bibitem{bel}
V. M. Belyaev, Phys. Lett. {\bf B254}, 153 (1991).
%
\bibitem{bgkp}
T. Bhattacharya, A. Gocksch, C. P. Korthals Altes, and R. D. Pisarski,
Phys. Rev. Lett., {\bf 66}, 998 (1991);
Nucl. Phys., {\bf B383}, 497 (1992);
A. Gocksch and R. D. Pisarski, Nucl. Phys. {\bf B402}, 657 (1993);
C. P. Korthals Altes,
hep-th/9310195, to appear in {\it Nucl. Phys. B}.
%
\bibitem{do}
V. Dixit and M. C. Ogilvie, Phys. Lett., {\bf B269}, 353 (1991).
%
\bibitem{ikr}
J. Ignatius, K. Kajantie, and K. Rummukainen, Phys. Rev. Lett.,
{\bf 68}, 737 (1992).
%
\bibitem{van}
V. M. Belyaev, I. I. Kogan, G. W. Semenoff, and N. Weiss,
Phys. Lett., {\bf B277}, 331 (1992);
W. Chen, M. I. Dobroliubov, and G. W. Semenoff,
Phys. Rev. {\bf D46}, R1223 (1992).
%
\bibitem{russians}
O. A. Borisenko, V. K. Petrov, G. M. Zinovjev,
Phys. Lett. {\bf B264}, 166 (1991);
V. V. Skalozub, Mod. Phys. Lett. {\bf A7}, 2895 (1992);
O. K. Kalashnikov, Phys. Lett. {\bf B302}, 453 (1993);
V. V. Skalozub, Trieste preprint IC/92/405, to appear in Phys. Rev.;
O. A. Borisenko, J. Boh\`a\v cik, V. V. Skalozub, (May, 1994) hep-ph/9405208;
O. K. Kalashnikov, Yukawa Inst. preprint YITP/K-1071 (April, 1994),
hep-ph/9405263.
%
\bibitem{smilga}
A. V. Smilga, Bern University preprint BUTP-93-03, (May, 1993) and
Santa Barbara preprint NSF-ITP-93-120, (Dec., 1993);
Univ. of Minn. preprint TPI-MINN-94-6-T, (Feb 1994),
hep-th/9402066;
I. I. Kogan, Princeton University preprint PUPT-1415, (Nov. 1993),
hep-th/9311164;
T. H. Hansson, H. B. Nielsen, and I. Zahed, USITP-94-09, hep-ph/9405234,
(May, 1994).
%
\bibitem{sintra}
C. Korthals Altes, K. Lee, and R. D. Pisarski, to appear in the
proceedings of the NATO workshop on ``Electroweak Physics and the
Early Universe,'' Sintra, Portugal (March, 1994);
and manuscript in preparation.
\bibitem{footnote1}
At higher order the relationship
between $A_0$ and the eigenvalues of $\Omega$ is more involved
\cite{bel,bgkp}.
Notice that we deal with all eigenvalues of the Wilson line; this
is in contrast to the Polyakov line, $=tr(\Omega)$, which is related
to the sum of the eigenvalues.
%
\bibitem{false}
S. Coleman, Phys. Rev., {\bf D15}, 292 (1977); C. Callan and S. Coleman,
Phys. Rev. {\bf D16}, 1762 (1977);
A. D. Linde, Phys. Lett. {\bf 100B}, 37 (1981);
A. D. Linde, Nucl. Phys. {\bf B216}, 421 (1983).
%
\bibitem{ew}
See the Proceedings of the NATO workshop on ``Electroweak Physics and the
Early Universe,'' Sintra, Portugal (March, 1994).
%
\end{references}
\end{document}